\documentclass[11pt,a4paper]{article}
\usepackage[makeroom]{cancel}
\usepackage{bm}
\usepackage{amsmath}
\usepackage{amsfonts}
\usepackage{amssymb}
\usepackage{graphicx}
\usepackage{authblk}
\usepackage{cite}
\usepackage[left=2.4cm,top=2.4cm,right=2.4cm,bottom=2.4cm,nohead]{geometry}
\DeclareGraphicsExtensions{.png,.jpg,.pdf}
\usepackage{epstopdf}
\usepackage{float}
\usepackage{fouriernc} 
\usepackage{enumitem}
\usepackage[utf8]{inputenc}
\usepackage[english]{babel}
\usepackage{mathtools}
\usepackage{amscd}
\usepackage{mathrsfs}
\usepackage{amsthm}
\usepackage{siunitx}
\usepackage{caption}
\usepackage{subcaption}
\setlist{nolistsep,leftmargin=*}

\usepackage{mhchem} 
\usepackage{cleveref} 
\usepackage{xcolor}
\usepackage{soul}
\usepackage{mdframed}
\usepackage{framed}
\usepackage{longtable}
\DeclareMathAlphabet{\mathpzc}{OT1}{pzc}{m}{it}

\title{General validity of the exponential law for the effect of concentration polarization in reverse osmosis in a stirred-cell geometry, including an activity correction for 1:1 salt solutions}

\makeatletter

\renewcommand\AB@authnote[1]{\textsuperscript{\normalfont#1}}

\makeatother

\author[1]{P.M. Biesheuvel,}
\author[2]{S.~Porada,}
\author[3,4]{I. Ryzhkov,}
\author[5]{M. Elimelech}

\affil[1]{Wetsus, European Centre of Excellence for Sustainable Water Technology,  
The~Netherlands.}
\affil[2]{Department of Process Engineering and Technology of Polymer and Carbon Materials, Wroclaw University of Science and Technology, Wyb. St. Wyspiańskiego 27, 50-370 Wrocław, Poland.}
\affil[3]{Institute of Computational Modeling SB RAS, Akademgorodok 50, 660036 Krasnoyarsk, Russia.}
\affil[4]{Siberian Federal University, Svobodny 79, 660041, Krasnoyarsk, Russia.}
\affil[5]{Department of Chemical and Environmental Engineering, Yale University, 
USA.}

\date{} 

\newcommand{\s}[1]{\mathrm{_{#1}}}

\begin{document}

\maketitle

\begin{abstract}

Reverse osmosis (RO) is a method to desalinate water, where water containing salts is pushed through a membrane while salt ions are rejected by the membrane. Very important in the theory of mass transport in RO is the concentration polarization (CP) layer, which develops on the upstream side of the membrane because of a combination of salt convection and diffusion. Because of the CP-effect, the salt concentration at the membrane surface is higher than in the channel, and this increases the osmotic pressure there, and thus transmembrane water flux is reduced (the osmotic pressure acts against water flux), while salt leakage through the membrane increases. So it is very important to understand and describe the CP-layer accurately. We analyze a one-dimensional geometry, which is of relevance for a typical lab-scale RO setup using small membrane coupons where the solution on the feed side of the membrane is stirred. For this geometry, the standard film layer approach is often used that assumes a stagnant film layer of a defined thickness, which however does not exist in reality. We set up a model without that assumption but including refreshment of solution because of the flow of water along the membrane due to stirring. We show that the `exponential law' for the CP-layer that is predicted by the the film model, also applies for this more accurate model. We further improve the model by including the activity coefficient of salt ions, as described by the Bjerrum theory that is based on ion-ion Coulombic interactions. We evaluate the original linearized Bjerrum theory as well as an extended Bjerrum equation that is valid up to 1.5 M salt concentration. We show how including this activity correction leads to a reduction of the diffusional driving force at high concentration, and thus the salt concentration at the membrane further increases. However, the effect can be easily included by reducing the CP-layer mass transfer coefficient by a fixed percentage. 

\end{abstract} 

\section*{Introduction}

Water can be desalinated with reverse osmosis (RO) membranes. By applying a pressure difference across the RO membrane, water flows through the sub-nanometer pores of a selective toplayer, while salts and other components in the water are largely retained. In this paper we describe theory for mixtures of neutral solutes and binary salts, with the results also relevant for multi-component salt mixtures. The theory also applies to membranes for nanofiltration (NF), which have larger pores and can work at lower pressure, but have a lower rejection of salts than in RO. 

Because ions are rejected, in front of the membrane, on the upstream high-pressure side, a concentration polarization (CP) layer forms, across which the salt concentration increases, from the concentration in bulk solution, to the value right next to the membrane. This layer is also called the diffusion boundary layer (DBL). Much experimental and theoretical work addresses salt concentration profiles in the CP layer, often including the two-dimensional geometry of a flow channel~\cite{Starov_1993,Zydney_1997,Chmiel_2006,Biesheuvel_Dykstra_2020,Pranic_2021,Biesheuvel_2022,Biesheuvel_2023,Johnston_2023}. Instead, in this work we focus on the CP-layer in a one-dimensional geometry, which is of relevance for desalination in a typical lab-scale setup for RO and NF, where the feedwater is stirred while being pressed through a small planar membrane coupon. Thus, in this geometry we have the same transport properties everywhere on the surface of the membrane coupon.

The classical model for the CP-layer is based on the geometry of a film layer, which assumes a stagnant layer of predefined thickness $\delta$, across which water and solutes flow, between a bulk phase on one side of this layer, and the membrane on the other side. This geometry leads to the `logaritmic law' or `exponential law', where the concentration at the membrane surface, $c\s{m}$, depends linearly on the concentration on the other side of the layer, i.e., in the bulk phase $c\s{f}$ (with index f for `feed'), and depends exponentially on the ratio of transmembrane water flux, $J\s{w}$, over CP-layer mass transfer coefficient, $k_\s{CP}$. In the first part of this paper, we show that also a more detailed model that does not use the film layer concept, but includes continuous refreshment of solvent and solutes in the region near the membrane, leads to the same exponential law, with only the relation between solute diffusion coefficient, \textit{D}, and mass transfer coefficient, $k_\s{CP}$, different from the film layer model.

In the second part we extend this model and include an activity correction for ions, which is the lowering of the chemical potential of an ion because of Coulombic attractions between anions and cations, an effect which can be accurately described by the Bjerrum equation. When this effect is included in the model for the CP-layer, the driving force for diffusion is reduced, and therefore the salt concentration at the membrane surface will further increase. Because of the activity correction the exponential law is no longer exact, but we find that in practice it is still highly accurate because we can include the activity effect by a fixed reduction in the CP-layer mass transfer coefficient. In practice, this coefficient is experimentally determined for given experimental conditions (defined by stirring rate, temperature, cell geometry), and thus this reduction will be automatically included in this measurement. Thus, though activity effects play a role in the transport of ions in solution, equations based on mass transfer coefficients, such as the exponential laws, remain valid, because the activity effect will be absorbed in a reduced mass transfer coefficient. 

\section*{Theory for concentration polarization for ideal solutions}

We first discuss the film layer concept for an ideal solution, which is a solution that follows the ideal gas law. Thus interactions between solutes are neglected, i.e., the activitity coefficient is unity for all solutes. The standard film layer model assumes a constant solute flux across a layer of finite thickness, $\delta$, and in this most simple model, the mass transfer coefficient, $k_\s{CP}$, is given by $k_\s{CP}\!=\!D/\delta$, with \textit{D} the diffusion coefficient. Results we present are for a single type of neutral solute, with concentration \textit{c}, but in case of mixtures of neutral solutes, the same equations apply for each component separately. The equations also apply when (next to a mixture of neutral solutes) the solution contains a binary salt, and then \textit{D} is to be interpreted as the harmonic mean diffusion coefficient, $D\s{hm}$, and \textit{c} is 
salt concentration. A binary salt solution has one cation and one anion, but the valencies can be different, as well as the diffusion coefficients of cation and anion. For instance, when only \ce{Ca^2+} and \ce{Cl-} are present, we have a binary salt solution. Reactions between ions and other solutes (e.g., salt pair formation, protonation reactions) are not considered.

\newpage

The flux of a solute $J\s{s}$ (unit mol/m\textsuperscript{2}/s) is a combination of diffusion and convection, and is given by  
\begin{equation}
J\s{s} = J\s{w} c - D \frac{\partial c}{\partial x} 
\label{eq_conv_diff}
\end{equation}
where $J\s{w}$ is the volumetric flux, or velocity, of solvent, here assumed to be water, with unit m/s, and \textit{x} a coordinate pointed towards the membrane, so $J\s{s}$ and $J\s{w}$ are both positive quantities. Both fluxes are superficial, i.e., defined per total unit cross-sectional area, irrespective of the presence of a porous structure in the flow channel. If there is such a porous material, the diffusion coefficient \textit{D} can be multiplied by (i.e., can include) a factor $\varepsilon$, which is the porosity, \textit{p}, divided by the tortuosity factor, $\boldsymbol{\tau}$, of that porous medium.  

In the standard film layer concept, it is assumed that this flux is constant across a layer of thickness $\delta$, i.e., steady state is assumed, and there are no flows entering or leaving `from the side', i.e., a 1D problem is solved. Then $J\s{s}$ and $J\s{w}$ are constant, and we can integrate across the CP-layer and arrive at
\begin{equation}
c_{\text{m}}  = \left(  c_{\text{f}} - \frac{J\s{s}}{J\s{w}} \right) \cdot  \exp\left(\frac{J\s{w}}{k_\s{CP}}\right)  + \frac{J\s{s}}{J\s{w}}
\label{eq_exp_1}
\end{equation}
where $k_\s{CP}=D/\delta$. Eq.~\eqref{eq_exp_1} can be used for each solute in a mixture of neutral solutes, and can also be used for the flow of a binary salt, with now \textit{c} the salt concentration, which often is denoted by $c_\infty$. Index m refers to a position right at the membrane surface, and is sometimes called the membrane concentration, while index f refers to the feed solution, i.e., outside the film layer in bulk solution. Eq.~\eqref{eq_exp_1} can also be written as function of the Péclet number, Pe, which is given by $\text{Pe} = \frac{J\s{w}}{k_\s{CP}}$. For certain conditions the concentration in the permeate, which is the water passing the membrane, is given by $c\s{p}=J\s{s}/J\s{w}$, which can then also be implemented in Eq.~\eqref{eq_exp_1}, resulting in the expression $\left(c\s{m}-c\s{p}\right)/\left(c\s{f}-c\s{p}\right)=\exp\left(\text{Pe}\right)$. The required condition, for this equation to be valid, is that on the permeate side there isn't already a volume of water of a different concentration, or water with another concentration flowing past the membrane. Instead, the concentration of the permeate must be determined solely by the local flow of solute and solvent through the membrane.

When the membrane retains the solutes perfectly, thus solute flux is zero, $J\s{s}\!=\!0$, then we obtain the simplified equation given by
\begin{equation}
c_{\text{m}}  = c_{\text{f}} \cdot  \exp\left(\frac{J\s{w}}{k_\s{CP}}\right) \, .
\label{eq_exp_2}
\end{equation}

Both these equations we call the `exponential law', with the first equation generally valid also with a non-zero solute flux through the membrane, while the second equation assumes perfect rejection of solutes. Both equations are based on the geometry of a thin stagnant layer with only diffusion and convection across its thickness, i.e., the film layer concept. They are very classical and are often used in calculations for reverse osmosis and nanofiltration. The question is whether these equations are also valid when a more realistic model is used for the flow of solvent and solutes in the region near the membrane. 

To address that question, we set up an improved model, where we include that stirring leads to the flow of solution along the membrane, which implies that solution at a certain distance from the membrane is swept away, and replaced there by fresh solution (from the bulk). This is the concept of `refreshment', familiar in mass transfer theory in chemical engineering. This approach can be used if we assume some level of turbulence, with local circulation (`eddies') involved in this mixing, or if we think more in terms of laminar flow paths oriented along the surface. In both cases the intensity of mixing will die out the closer we approach the membrane, with the eddies smaller and less vigorous, to disappear completely right at the membrane, while for the assumption of parallel flow lines, the velocity of fluid along these lines is large at some distance away, but decreases when we approach the membrane, and finally becomes zero at the membrane because of the zero slip condition. Both a high level  of stirring and high flow velocities lead to fast refreshment, i.e., a high frequency by which packages of solvent and solutes in the layer near the membrane are swept away and replaced by fresh solution. This frequency is high far from the membrane and decreases to zero right at the membrane: ultimately, in the very last nm before the membrane, we only have diffusion and convection in the direction right into the membrane, and the refreshment effect is gone. So with a certain frequency, $f$, dependent on distance from the membrane, some solution in the transport layer is swept away and replaced, or refreshed, with bulk solution. This replacement frequency, $f$, increases linearly with distance from the membrane, starting  at zero at the membrane surface. The gradient (change with position) of this frequency, $g$, is a measure of the intensity of stirring, of refreshment, i.e., a faster increase in frequency means there is more intense stirring, corresponding to a thinner film layer in the classical approach. We can also interpret \textit{f} as a velocity along the surface, $v_\parallel$, divided by a distance along the membrane, $\ell$, for a volume of fluid to travel before being mixed up.

We solve this model by setting up a one-dimensional solute mass balance and assume steady state (no time dependencies). We include convection, diffusion, and refreshment, and then arrive at
\begin{equation}
J\s{w} \frac{\partial c}{\partial \widetilde{x}} + D \frac{\partial^2 c}{\partial \widetilde{x}^2} + f \, \left( c\s{f} - c\right) = 0
\label{eq_mass_balance_1}
\end{equation}
where $\widetilde{x}=-x$ is a coordinate pointing away from the membrane ($\widetilde{x}\!=\! 0$ at the membrane). The velocity of water towards the membrane, $J\s{w}$ (which we use as a positive number), is assumed to be independent of coordinate $\widetilde{x}$. In a stirred cell, $J\s{w}$ is constant because there are no pressure or concentration gradients along the surface, i.e., everywhere on the coupon surface we have the same conditions. So it would be impossible for water and solutes arriving in the region near the membrane to go sideways in any particular direction, as all directions are the same. Water and solutes have no other option than to flow straight to the membrane. There is the additional effect of refreshment (mixing), but this also has no directionality, i.e., mathematically it functions as a source/sink-term, and will not change that 
$J\s{w}$ must point straight towards the membrane, and as a consequence can be treated as a constant. 

We introduce the concentration difference $y=c-c\s{f}$, as well as a non-dimensional coordinate $\xi= \widetilde{x} / \sqrt[3]{D/g}$ with $g$ the above-discussed gradient in injection frequency (unit (m.s)\textsuperscript{-1}), related to $f$ by $f = g  \widetilde{x}$. This is similar to the two-dimensional approach of this problem, where $g$ defined above is replaced by the shear rate $\gamma$ (unit s\textsuperscript{-1}) divided by membrane length, \textit{L}.

Implementing these conversions, Eq.~\eqref{eq_mass_balance_1} becomes
\begin{equation}
\alpha \cdot \text{Pe} \cdot y' + y'' - \xi \cdot y = 0
\label{eq_mass_balance_2}
\end{equation}
where $y'$ is shorthand for $\partial y / \partial \xi$, and $y''$ for $\partial^2 y / \partial \xi^2$. The reason for introducing the factor \mbox{$\alpha \! = \! 0.729011 \dots $} will be apparent in a moment, when we analyze Eq.~\eqref{eq_mass_balance_2} in the absence of convection ($J\s{w}\!=\!0$, thus $\text{Pe}\!=\!0$), and in that case Eq.~\eqref{eq_mass_balance_2} becomes the Airy equation. In the absence of convection ($\text{Pe} \! = \! 0$), Eq.~\eqref{eq_mass_balance_2} results in the classical expression $J\s{s} = k_\s{CP} \, \left(c\s{f} - c\s{m} \right)$ for the flux of solute (or salt) through the interface. Thus, for $\text{Pe}\!=\!0$, the same as for the standard film layer model, we have a linear relationship between the concentrations on the two ends of a mass transfer zone, and the resulting solute flux. The only difference is that in this case the mass transfer coefficient is given by $k_\s{CP} = \alpha \cdot \sqrt[3]{D^2 \, g}$. Interestingly, according to this expression for $k_\s{CP}$, a 10$\times$ larger intensity of stirring (assuming that would lead to a 10$\times$ higher $g$), only leads to a roughly 2.1$\times$ larger rate of mass transfer: stirring is good, but too much stirring will not be effective.

In this model which includes refreshment, which is the third term in Eq.~\eqref{eq_mass_balance_2}, solute concentration does not linearly change across the film layer, with a sudden start on the outside (where $\widetilde{x}=\delta$) as in the standard film model, but it now gradually starts to increase from some distance away, and this profile of concentration versus position becomes increasingly steep the closer we approach the membrane. Thus, in a transport model with refreshment, we have a more realistic description of the concentration profile than in the standard film model, but the mathematical form of solute flux, $J\s{s}$, as function of the concentration difference, $c\s{f}  - c\s{m}$, is exactly the same, only with a different dependency of $k_\s{CP}$ on $D$. 

Next, we consider the case that convection is included, i.e., $\text{Pe} \neq 0$, and study whether the exponential law, that was derived for the standard film model, also applies in this more advanced geometry. To evaluate Eq.~\eqref{eq_mass_balance_2} including convection ($\text{Pe} \neq 0$), we use a numerical procedure where we solve Eq.~\eqref{eq_mass_balance_2} for a certain value of Pe, and for a certain concentration at the membrane, i.e., $y\!=\! y_0$ at $\xi_0 \! = \! 0$ ($y_0 \! = \!  c\s{m} - c\s{f}$), to find the gradient at the surface from the numerical calculation, $y'_\text{num,0}$, such that in the calculation far from the surface we simultaneously attain $y \! = \! 0$ and $y' \! = \! 0$, which typically occurs at $\xi$~between~2~and~4. This gradient relates to solute flux through the membrane according to
\begin{equation*}
J\s{s}- c\s{m} J\s{w} = D \, \left. \frac{\partial c}{\partial \widetilde{x}}\right|_0 = \sqrt[3]{D^2 g} \cdot \left. \frac{\partial c}{\partial \xi}\right|_0  =  {k_\s{CP} }  \cdot \frac { y'_\text{num,0} }{ \alpha}
\end{equation*}
with $c\s{m}$ the concentration at the membrane, where $\xi\!=\!0$. The exponential law, Eq.~\eqref{eq_exp_1}, can be rewritten to the similar form 
\begin{equation*}
J\s{s} - c\s{m} J\s{w} = - k_\s{CP} \cdot  \frac{y_0 \, \text{Pe}}{1-e^{-\text{Pe}}}   
\end{equation*}
and thus, to check whether the full model of Eq.~\eqref{eq_mass_balance_2} agrees with the exponential law, we compare 
\begin{equation}
  \frac{ y'_\text{num,0} }{\alpha}  \leftrightarrow  - \frac{y_0 \, \text{Pe}}{1-e^{-\text{Pe}}}  
\end{equation}
and see how close these two groups are. Up to $\text{Pe}\!=\! 6$, which is the maximum that we tested (and which is significantly beyond Pe-numbers that apply to RO experiments, which typically are below $\text{Pe}\!  = \! 1$), the difference is always less than 2\%. In addition, below $\text{Pe}\!=\!1.1$ and above $\text{Pe}\!=\!5.5$, we find the difference to be no more than 1\%. Thus we can conclude that Eq.~\eqref{eq_exp_1} not only applies to the film layer model, but also for the more accurate model including refreshment that we analysed in this section. Thus Eq.~\eqref{eq_exp_2}, which is a specific version of Eq.~\eqref{eq_exp_1}, valid when the membrane rejects the solutes perfectly, also applies for the advanced model based on Eq.~\eqref{eq_mass_balance_1} that includes refreshment.

So our results show that the exponential laws, Eqs.~\eqref{eq_exp_1} and~\eqref{eq_exp_2}, are accurate for reverse osmosis and nanofiltration from a stirred solution of neutral solutes that follow ideal `gas law' statistics. For an ideal binary salt solution, this is also the case, except for a deviation because ionic solutions are not ideal (in the sense of following ideal gas statistics). Instead, the activity coefficient of ions is less than unity, decreasing with salt concentration. This will effectively lead to a slower diffusion of ions back from the membrane into solution, and will therefore lead to higher concentrations there, beyond what the exponential law predicts.

\section*{Theory for non-ideal solutions, including activity corrections~-~I}

For a 1:1 salt, we can include an activity correction for ions in solution based on the Bjerrum theory. If a concentration gradient develops in a region, the activity correction leads to an additional force on ions, resulting in a reduced tendency to diffuse to low concentrations. This force can be combined with that for regular diffusion and then we can still use Eq.~\eqref{eq_conv_diff} but with a diffusion coefficient that 
has an activity correction. The Bjerrum theory describes how the activity correction, $\ln\gamma$, is proportional to the cube root of salt concentration and for a 1:1 salt such as NaCl, this model is correct up to a salt concentration of $\sim 200$~mM~\cite{Biesheuvel_2020}. This regular, or linear, Bjerrum equation is used in this section, while in the next section we use an extended Bjerrum equation for $\ln\gamma$ which is valid up to a salt concentration of 1.5~M. In the regular Bjerrum theory, the constant diffusion coefficient \textit{D} is replaced by the concentration-dependent function
\begin{equation}
D = D_0 \cdot \left( 1- \vartheta \sqrt[3]{c}  \right)
\label{eq_bjerrum_diff_coeff}
\end{equation}
where $D_0$ is the diffusion coefficient in the dilute limit. In Eq.~\eqref{eq_bjerrum_diff_coeff} we introduce the factor $\vartheta$ which is $\vartheta \sim 0.0202$~m/mol\textsuperscript{1/3} for a 1:1 salt at room temperature~\cite{Biesheuvel_2020}. This equation accurately describes data (compared in the range up to 200~mM) for the rate of diffusion as function of salt concentration for several types of salts~\cite{Biesheuvel_2020}. Eq.~\eqref{eq_bjerrum_diff_coeff} predicts a 12\% reduction in the rate of diffusion at $c\!=\!200$~mM.

With this effect included, Eq.~\eqref{eq_mass_balance_1} becomes
\begin{equation}
J\s{w} \cdot \frac{\partial c}{\partial \widetilde{x}} + D_0 \cdot \frac{\partial}{\partial \widetilde{x}}\left( \left(1-\vartheta \sqrt[3]{c}  \right) \, \frac{\partial c}{\partial \widetilde{x}}  \right) + f \cdot \left( c\s{f} - c\right) = 0
\label{eq_mass_balance_3}
\end{equation}
which we can rewrite to
\begin{equation}
\alpha \cdot \text{Pe} \cdot \frac{\partial c}{\partial \xi} +  \left(1-\vartheta {c}^{1/3}  \right) \cdot \frac{\partial^2 c}{\partial \xi^2} -  \tfrac{1}{3} \cdot \vartheta \cdot c^{-2/3} \cdot \left( \frac{\partial c}{\partial \xi} \right)^2  +  \xi \cdot \left( c\s{f} - c\right) = 0
\label{eq_mass_balance_5}
\end{equation}
where the definitions of Pe, $\xi$, and $k_\s{CP}$ are now based on $D_0$, not $D$. Here, in Eq.~\eqref{eq_mass_balance_5} we do not make use of the factor \textit{y} because it would not simplify the problem. So this problem is no longer invariant with concentration \textit{c}, as it was in the theory of the last section that did not include an activity correction. There, a 10- or 100- or 1000-fold increase in concentration did not impact the assessment on the accuracy of the exponential law. That is different when an activity correction is included, where we do have a direct dependency on concentration, \textit{c}. Eq.~\eqref{eq_mass_balance_5} reduces to Eq.~\eqref{eq_mass_balance_2} for $\vartheta \! \to \! \!0$ or $c \! \to \! 0$.  

We make a calculation for a perfect membrane (i.e., at $\xi \!= \! 0$ the salt flux is $J\s{s}\!=\! 0$), at Pe-values up to $\text{Pe} \! = \! 0.8$. In practice, this maximum value of Pe, which describes a high water flux, is only reached when the feed salt concentration is low, see for instance Fig.~3C in ref.~\cite{Biesheuvel_2023}. We compare the predicted membrane concentration, $c\s{m}$, with and without the activity effect, in the latter case by the use of the exponential law, Eq.~\eqref{eq_exp_2}. The zero salt flux at the membrane surface leads to 
\begin{equation}
D_0 \cdot \left( 1 - \vartheta c\s{m}^{1/3} \right) \cdot \left.\frac{\partial c}{\partial \widetilde{x}}\right|_{\widetilde{x}=0} + J\s{w} \cdot c\s{m} = 0
\end{equation}
which can be rewritten to the boundary condition
\begin{equation}
\ \left.\frac{\partial c}{\partial \xi}\right|_{\xi=0} = - \frac{\alpha \cdot \text{Pe} \cdot c\s{m}}{1 - \vartheta \cdot c\s{m}^{1/3}}  
\end{equation}
which we will use in the numerical calculation. 

Calculation results with this numerical model are that the salt concentration at the membrane, $c\s{m}$, is now higher than in the original model without the activity effect. For instance, at $\text{Pe}_0\!=\!0.8$ and $c\s{f}\!=\!500$~mM, we now have $c\s{m}\!=\!1245$~mM, which according to the exponential law would have been $c\s{m}\!=\!1113$~mM, and thus we have an increase of 12\% in concentration. Here index `0' refers to evaluation of the Pe-number based on $D_0$, i.e., without the correction we will introduce further on.

It now turns out that we can describe all calculation output quite accurately when we do not use $k_\s{CP}$ based on $D_0$, but reduce $k_\s{CP}$ by a fixed percentage. Thus, it is not needed to make $k_\s{CP}$ a function of $c\s{f}$, or 
of some average of $c\s{f}$ and $c\s{m}$, or introduce a dependency on the water velocity. 
Instead, the only adjustment is simply to reduce $k_\s{CP}$ by a fixed percentage, the same for all conditions of water velocity (thus Pe-number) and feed concentration. Thus, also with the activity effect included, the entire problem is described by a fixed value of $k_\s{CP}$. This is then the same as how the film model is used without activity effects, with typically $k_\s{CP}$ derived from fitting a model for water and salt flux to data. So even when we include activity effects, the procedure to analyze data and then to fit a single value of $k_\s{CP}$ to all data of water and salt flux, 
remains valid. And the accuracy of the exponential law remains the same. 

The percentage by which $k_\s{CP}$ is reduced relative to $k_\s{CP,0}$, the latter based on $D_0$, will depend on the activity model. For the Bjerrum equation, Eq.~\eqref{eq_bjerrum_diff_coeff}, with the present choice of $\vartheta = 0.0202$~m/mol\textsuperscript{1/3} (a value which depends on temperature and solvent type), we find that a 10\% reduction is optimal. Thus, if an estimate of $k_\s{CP}$ based on $D_0$ would have been $k_\s{CP,0}\!=\!100$~L/m\textsuperscript{2}/h (LMH), the output of the model that includes an activity correction, can be reproduced by the exponential laws of Eqs.~\eqref{eq_exp_1} and~\eqref{eq_exp_2} when we multiply $k_\s{CP,0}$ by 0.9, and thus the effective value of $k_{\s{CP,eff}}$ to be used in Eqs.~\eqref{eq_exp_1} and~\eqref{eq_exp_2} is $k_{\s{CP,eff}}=0.9 \times 100 =90$~LMH. With this correction, for the numerical calculation discussed before, at $\text{Pe}_0\!=\!0.8$, the exponential law of Eq.~\eqref{eq_exp_2} (modified because $k_\s{CP}$ is reduced by 10\%) now predicts $c\s{m} \! = \! 1216$~mM, instead of 1113~mM, which is only a 2.3\% underestimate of the exact value of 1245~mM. Now, these high salt concentrations and high Pe-numbers are not generally encountered simultaneously in RO experiments, see Fig.~3C in ref.~\cite{Biesheuvel_2023}. So we next analyze lower Pe-values and lower concentrations. At $c\s{f}\!=\! 500$~mM, the deviation of 2.3\% at $\text{Pe}_0\!=\!0.8$ drops to 1.6\% at $\text{Pe}_0\!=\!0.6$, to 0.9\% at $\text{Pe}_0\!=\!0.4$, and to 0.3\% at $\text{Pe}_0 \! = \! 0.2$. At $c\s{f}\!=\!400$~mM the deviation is 1.3\% at $\text{Pe}_0\!=\!0.8$, while at $\text{Pe}_0\!=\!0.6$ and below, it is less than 1\%. For $c\s{f}\!=\!300$~mM the deviation is less than 1~mM at all Pe-values considered, while at still lower salt concentrations this deviation is 5--8 mM at $\text{Pe}_0\!=\!0.8$, around 4~mM at $\text{Pe}_0\!=\!0.6$, and less than 2~mM at $\text{Pe}_0\!=\!0.3$. 

Thus, we can conclude that the activity effect for salt solutions, which influences the rate of salt diffusion, and thus the salt concentration at the membrane, does not have to be explicitly considered, but the effect of activity can be accurately incorporated by making use of a lower mass transfer coefficient, with the reduction independent of water flux or salt concentration. Because the standard procedure in experimental work is to fit a value of the mass transfer coefficient to actual data, there is no need to explicitly account for the activity correction in a mass transfer model for the CP layer, at least not in the case considered, which is for a simple 1:1 salt. Thus, in a one-dimensional geometry, which describes a stirred cell, the exponential laws, Eqs.~\eqref{eq_exp_1} and~\eqref{eq_exp_2}, can be used with confidence for neutral solutes as well as for 1:1 binary salt solutions.

\section*{Theory for non-ideal solutions, including activity corrections~-~II}

A problem in the analysis of the last section is that an expression was used for the activity correction, $\ln\gamma$, that is only valid up to approx.~200~mM. Beyond that salt concentration, $\ln\gamma$ does not decrease as fast as that expression predicts, and $\ln\gamma$ even starts to increase again; for NaCl, the minimum is found at approx.~1.0~M, where $\ln\gamma$ is $\sim \! -0.45$. However, at that salt concentration, the linear Bjerrum theory of the previous section predicts $\ln\gamma = - 0.60$, which is too low. An improved expression was derived for $\ln\gamma$ in ref.~\cite{Biesheuvel_2020} (Appendix I there) which we here write as
\begin{equation}
\ln\gamma_\pm = - b \cdot c^{1/3} - \tfrac{1}{4} \cdot b^2 \cdot c^{2/3} + 6 \cdot b^3 \cdot q \cdot c
\label{eq_extended_bjerrum_lng}
\end{equation}
where \textit{b} is a factor that for a 1:1 salt at room temperature is $b= 0.0605\text{~mM}^{-1/3}$, and where \textit{q} is a factor that relates to the average radius of the ions, $\left\langle a \right\rangle$. 
We find that data for NaCl up to 1.5~M (and accepting a small error, even up to 2.0~M) are well described using $q\!=\!0.19$, see also ref.~\cite{Biesheuvel_2022_B}. 


Using this extended Bjerrum equation, the diffusion coefficient can be expressed as
\begin{equation}
D = D_0 \cdot \left( 1 - \vartheta_1 \, c^{1/3} - \vartheta_2^2 \, c^{2/3} + \vartheta_3^3 \, c \right)
\label{eq_EXT_bjerrum_diff_coeff}
\end{equation}
where $\vartheta_1 = 0.0202 $, $\vartheta_2 = 0.0247$, and $\vartheta_3 = 0.0632$, all with unit~$\text{mM}^{-1/3}$~($\vartheta_3$ is based on $q\!=\!0.19$).
Eq.~\eqref{eq_mass_balance_3} is now modified to
\begin{equation}
J\s{w} \cdot \frac{\partial c}{\partial \widetilde{x}} + D_0 \cdot \frac{\partial}{\partial \widetilde{x}}\left( \left(1-\vartheta_1 c^{1/3} - \vartheta_2^2 c^{2/3} + \vartheta_3^3 c  \right) \cdot \frac{\partial c}{\partial \widetilde{x}}  \right) + f \cdot \left( c\s{f} - c\right) = 0
\label{eq_mass_balance_6}
\end{equation}
which we can rewrite to
\begin{equation}
\alpha \cdot \text{Pe} \cdot \frac{\partial c}{\partial \xi} +  \left(1-\vartheta_1 {c}^{1/3} - \vartheta_2^2 c^{2/3} + \vartheta_3^3 c  \right) \cdot \frac{\partial^2 c}{\partial \xi^2} - \left( \tfrac{1}{3} \vartheta_1 c^{-2/3} + \tfrac{2}{3} \vartheta_2^2 c^{-1/3} - \vartheta_3^3  \right) \cdot \left( \frac{\partial c}{\partial \xi} \right)^2  +  \xi \cdot \left( c\s{f} - c\right) = 0  \, .
\label{eq_mass_balance_7}
\end{equation}
For a perfect membrane, the zero salt flux boundary condition now becomes 
%
\begin{equation}
\ \left.\frac{\partial c}{\partial \xi}\right|_{\xi=0} = - \frac{\alpha \cdot \text{Pe} \cdot c\s{m}}{1 - \vartheta_1 \cdot c\s{m}^{1/3} - \vartheta_2^2 \cdot c\s{m}^{2/3} + \vartheta_3^3 \cdot c\s{m}}  \, .
\end{equation}

Using this model we make similar calculations as in the previous section, and we now find that the salt concentration at the membrane, $c\s{m}$, is still higher than in the ideal case, but the deviation is very minor now. At $\text{Pe}_0\!=\!0.8$ and $c\s{f}\!=\!500$~mM, we now find $c\s{m}\!=\!1124$~mM, which is very close to $c\s{m}$ predicted the ideal exponential law which is $c\s{m}\!=\!1113$~mM.  
So the extended Bjerrum theory leads to a concentration at the membrane only 11 mM different from that predicted by the ideal exponential law, which is less than 1\% different, while there was a difference of 132~mM in the earlier analysis. 
At lower feed salt concentrations (400, 300 and 200~mM), with the same $\text{Pe}_0\!=\!0.8$, the ideal law underestimates the calculation result using the extended Bjerrum equation by $\sim\!20$~mM at each of these salt concentrations, which is $\sim \!2 \!  - \! 4  \times$ less than the deviation from the linear Bjerrum equation. This small difference can be further diminished by the method of reducing $k_\s{CP}$ in the exponental law, and now a reduction of 2.5\% in $k_\s{CP}$ is sufficient to bring the remaining difference (between exponential law, and full calculation using the extended Bjerrum equation) to $\sim \! 1 \%$ for $c\s{f}$ between 300 and 500 mM, and $\sim \! 2 \%$ for 200 mM feed concentration. So again the conclusion is that a small reduction in $k_\s{CP}$ is sufficient to make the ideal exponential laws fit with full numerical theories that include ion activity corrections. 

\section*{Conclusions}

The exponential law describes concentration polarization in reverse osmosis and nanofiltration in a one-dimensional geometry, such as when the feedwater is a stirred batch volume that is pushed through a flat membrane coupon. It was originally derived based on the simple film layer geometry, and the assumption that solutes behave thermodynamically ideal. We develop an advanced flow model that does not use the film layer concept, but includes solute refreshment in the region near the membrane, and for a 1:1 salt solution we include an ion activity effect using the Bjerrum theory. We find that the original exponential law for the CP-layer is also valid for the advanced flow model, and also when an activity correction is considered. We find that the activity correction does not need to be included explicitly because it can be incorporated in the CP-layer mass transfer coefficient, which in most studies is obtained from simultaneously fitting solvent and solute flux from a membrane model to experimental data, and then the activity correction is automatically included.

\end{document}